\documentclass[showpacs,preprintnumbers,amsmath,amssymb]{revtex4}
\usepackage{graphicx}
  \usepackage{dcolumn}
  \usepackage{bm}
   \usepackage{amsmath,amssymb}

\begin{document}
  \title{Chiral
  Thirring-Wess Model}
\author{Anisur Rahaman}
\email{anisur.rahman@saha.ac.in ;  manisurn@gmail.com}
\affiliation{Hooghly Mohsin College, Chinsurah, Hooghly - 712101,
West Bengal, India}

 \begin{abstract}
  The vector type of interaction of the Thirring-Wess model was replaced by the chiral
   type and a new model was presented which was termed  as chiral Thirring-Wess model in
    \cite{THAR}. The model was studied there with a Faddeevian class of regularization.
     Few ambiguity parameters were allowed there with the apprehension that unitarity
     might be threatened like the chiral generation of the Schwinger model. In the present
      work it has been shown that no counter term containing the regularization ambiguity
      is needed for this model to be physically sensible.  So the chiral Thirring-Wess
      model is studied here without the presence of any ambiguity parameter and it has
      been found that the model not only remains exactly solvable but also does not lose
      the unitarity like the chiral generation of the Schwinger model. The phase space
      structure and the theoretical spectrum of this new model have been determined in
      the present scenario. The theoretical spectrum is found to contain a massive boson
       with ambiguity free mass and a massless boson.

\end{abstract}
  \maketitle

 \section{Introduction}
  In lower dimensional field theory Thirring and Wess jointly presented an
  interesting and exactly solvable model \cite{THIR} few years after the presentation
  of the Schwinger model \cite{SCH}. The nature of interaction between the matter and
   gauge field in these two models were identical and both the models were
   capable of describing generation of mass and had been studied over the
    years not only because of their  surprising exactly solvable nature but
    also for their ability to describe mass generation, confinement as well as
    de-confinement aspect of fermions etc.,
     \cite{THIR, SCH, LOW, COL, CAS, AG, ARS, THAR}. In the Schwinger model
     it was found that photon acquired mass via a kind of dynamical symmetry
      breaking keeping the gauge symmetry of the model intact. Recently, it
       was shown by us that  a novel back ground interaction emerged out
        when an attempt was made  to study this model in the non commutating
        space time setting \cite{APR, APR1, SADO}.

Though the joint analysis of Thirring and Wess \cite{THIR}
revealed that mass generation of photon can also be explained by
the model coined by themselves nevertheless gauge symmetric
structure  was not possible to maintain in this model to start
with \cite{THIR}. This model was named after Thirring and Wess and
it was known as Thirring-Wess model. Recently, an attempt has been
made in \cite{BELV}, for systematic functional integral
bosonization of this mode.

After few years of presentation of the Thirring-Wess model, chiral
generation of Schwinger model was perused  in \cite{HAG},  however
the model remaind less attractive over a long period of time
because of its inability to provide a unitary field theoretical
description.  But it attracted huge attentions and gradually
acquired a significant position in lower dimensional field theory
after the removal of the non-unitary problem  by Jackiw and
Rajaraman \cite{JR} taking into account anomaly into
consideration. Anomaly played interesting as well as surprising
role for this model to place it in a well acceptable position. The
welcome  entry of the anomaly and the suitable exploitation of the
ambiguity involved therein made Jackiw-Rajaraman version of Chiral
Schwinger model  \cite{JR, ROT1, ROT2, KH, MIYAKE0, MIYAKE} along
with the other independent regularized version of that model
\cite{PM, MG, SUBIR0, SUBIR} so popular in lower dimensional field
theory regime.

 Like the Schwinger model the Thirring-Wess model also describes an
  interacting theory of massless fermion  with Abelian  gauge field
   in two dimension and the nature of interaction in both the models
   are identical ( vector like). The difference lies only in the gauge
   field sector. In the Schwinger model the kinematics
   of the gauge field enters through the  Maxwell type lagrangian,
   however the Thirring-Wess model can be considered  as a study of
   QED, viz., Schwinger model \cite{SCH, LOW} replacing Maxwell's
   field by Proca field and that very replacement breaks the gauge
    symmetry of the model at the classical level. However,
     no  constraint related to inconsistency  stands  as a hindrance
      against the way of Thirring-Wess model \cite{THIR} to mark it as
       a sound field theoretical model.

It is true that the so called non-confining Schwinger model
\cite{AR, AR1,ARN}  is a structurally equivalent gauge
non-invariant model to the Thirring-Wess mode,  but there lies a
crucial difference between these two.  In the Thirring-Wess  mode
the masslike  term for the gauge field entered at the classical
level  however in the  so called non-confining Schwinger model the
same type of masslike term gets involved through the one loop
correction which contains an ambiguity parameter like
Jackiw-Rajaraman version of  Chiral Schwinger model.

An attempt was made in \cite{THAR} in order to have the chiral
  generation of the Thirring-Wess mode in the similar way the
   chiral generation  was made  for the Schwinger model in \cite{HAG}.
   The chiral generated version however had to be  studied there
    with a Faddeevian class of regularization and few ambiguity parameter
    were allowed to enter during the process of removal of divergence of the
    fermionic determinant in course of going over to its bosonized version with an
     apprehension that it might suffer from the unwanted  non unitarity
     problem like the chiral generation of the Schwinge model \cite{HAG}.

 Let us now explore a bit related to the necessity of the allowance of
 ambiguity parameter in the QED and chiral QED. It is known that the vector
  Schwinger model remained unitary and exactly solvable in absence of any
  ambiguity parameter, particularly when its  bosonized version is concerned,
   however
  the chiral generation of this model \cite{HAG} faced a severe non-unitary
   problem. The Thirring-Wess model in its usual version was free from
    non-unitary problem and was exactly solvable even in the absence of
     any ambiguity parameter in its bosonized version too.  A question,
     therefore, automatically generates whether the chiral generation of
      Thirring-Wess model can provide a physically sensible theory even
       in the absence of any ambiguity parameter like its ancestor where
        the basic interaction was vector like? The search for the answer
         to this question leads to an interesting investigations which we
          would like to explore in this paper.  To investigate it in detail
          an attempt has been made here to investigate the chiral Thirring-Wess
          model without any counter term containing ambiguity parameter. One
           may however be free to allow ambiguity involved counter term in
           this situation too as long as it remains physically sensible. When
           the ambiguity involved one loop correction is allowed to enter that
           also suggests another direction of investigation. It suggests how the
            theoretical spectrum gets modified when this one loop correction is
            allowed. A special type of ambiguity involved counter term is
             allowed in our analysis which resembles the
             Jackiw-Rajaraman's counter term for the chiral
             Scgwinger model
              and a
              pedagogical illustrations has been made to see the role of
               ambiguity parameter towards the modification of the theoretical spectrum.

 The paper is organized as follows. Sec. II, contains the chiral
  generation of Thirring-Wess model.  In Sec. III. , the model has
  been  analyzed using lagrangian formulation and free field solution
   is obtained. In Sec. IV the phase space structure of the mode is
    determined using Dirac method of quantization of constrained
     system.  Sec. V. contains the concluding remarks.

\section{Chiral generation of the Thirring-Wess model}
 The Chiral generation of the Thirring-Wess model was attempted in our previous work
 \cite{THAR} by the
 following generating functional
 \begin{equation}
 Z[A] = \int d\psi d\bar\psi e^{\int d^2x{\cal L}_f}
 \end{equation}
 with
 \begin{eqnarray}
 {\cal L}_f &=& \bar\psi\gamma^\mu[i\partial_\mu + e\sqrt\pi
 A_\mu(1-\gamma_5)]\psi \nonumber \\ &=& \bar\psi_R\gamma^\mu
 i\partial_\mu\psi_R +\bar\psi_L \gamma^\mu(i\partial_\mu +
 2e\sqrt\pi A_\mu)\psi. \label{LF}
  \end{eqnarray}
 Here dynamics of the $A_\mu$ field was governed by he Proca field
 and the lagrangian of which was given by
 \begin{equation}
 {\cal L}_{Praca}= \frac{1}{4} F_{\mu\nu}F^{\mu\nu} +\frac{1}{2}m^2
 A_\mu A^\mu. \label{PROCA}
 \end{equation}
 It is known that the vector and axial vector current that couple with the gauge
 field are defined by
  $J_\mu = \bar\psi
 \gamma_\mu \psi$ and $J^5_\mu = \bar\psi
 \gamma_\mu \gamma_5\psi$ respectively. In order to make chiral generation of the usual
  Thirring-Wess model we
 replaced the vector type of interaction $\bar\psi
 \gamma_\mu \psi A^{\mu}$ by the chiral type
 $\bar\psi\gamma_\mu(1-
 \gamma_{5})\psi A^{\mu}$.  It was mentioned there  that the right handed
  fermion remained uncoupled
 for this type of chiral interaction whereas the left handed fermion got
  coupled with the gauge field.
  As a result, integration over this right
 handed part led to a field independent counter part which was
 absorbed within the normalization. However, the integration over the left handed
 fermion was not like that. The computation of  the integration of
 the left handed part is much involved and plays crucial role to
 make the model sensible as well as  physically acceptable in the
 lower dimensional field theoretical regime. After integrating out
 of the lefthanded part the generating functional in a general
 form
 can be written down as
 \begin{eqnarray}
 Z[A] &=& \int d\psi_L d\bar\psi_L\bar\psi_L
 \gamma^\mu(i\partial_\mu + 2e\sqrt\pi
 A_\mu)\psi_L  \nonumber \\
 &=&exp\frac{ie^2}{2}\int d^2x A_\mu[M_{\mu\nu} - (\partial^\mu
 +\tilde\partial^\mu) \frac{1}{\Box} (\partial^\nu +
 \tilde\partial^\nu)]A_\nu. \label{GEN}
 \end{eqnarray}
Here $M_{\mu\nu}$ contains the one loop correction effect needed
to remove the divergence of the fermionic determinant. Some
regularization ambiguity also enters there during the course of
removing the divergence. In general $M_{\mu\nu}$ can take any form
however the maintenance of unitarity and Lorentz invariance put
different constraints and allows some suitable structure. One
admissible structure was considered in \cite{THAR}. In the present
context we are interested to study chiral generation of the
Thirring-Wess model with out any ambiguity involved term to
investigate whether it can provide any physically sensible theory.
It is known that in case of the chiral generation of the Schwinger
model \cite{HAG} it was not possible to have a sensible theory
without the help of any ambiguity parameter.

The above generating functional (\ref{GEN}) can be expressed in
terms of some auxiliary field and when it is done so
  in terms of the an auxiliary
 field $\phi(x)$ it turns out to the following
 \begin{equation}
 Z[A] = \int d\phi e^{i\int d^2x {\cal L}_B},
 \end{equation}
 with
 \begin{equation}
 {\cal L}_B = \frac{1}{2} (\partial_\mu\phi)(\partial^\mu\phi) +
 e(g^{\mu\nu} - \epsilon^{\mu\nu)}\partial_\nu\phi A_\mu +
 \frac{1}{2}e^2
 M_{\mu\nu} A_\mu A^\mu.
 \end{equation}
Since we are motivated here to study the chiral generation of
Thirring-Wess model with out any ambiguity parameter the effect of
$M_{\mu\nu}$ is ignored here and the total lagrangian density,
i.e., the bosonized lagrangian density along with the Proca
 background now reads
\begin{eqnarray}
{\cal L} &=& {\cal L}_B + {\cal L}_{Praca}, \nonumber \\
 &=& \frac{1}{2}(\partial_\mu\phi)(\partial^\mu\phi) + e(g^{\mu\nu}
 - \epsilon^{\mu\nu)}\partial_\nu\phi A_\mu
 - \frac{1}{4} F_{\mu\nu}F^{\mu\nu} +\frac{1}{2}m^2 A_\mu A^\mu, \nonumber \\
 &=& \frac{1}{2}(\dot\phi^2 - \phi'^2) + e(\dot\phi + \phi')(A_0
 - A_1)
 + \frac{1}{2} ({\dot A_1}^2-{A'_0}^2) +\frac{1}{2}m^2( A_0^2 - A_1^2)
 \label{PLAG}.\end{eqnarray}
We have chosen $\epsilon^{01} = +1$. Let us recall that when this
type of chiral generalization was attempted for  the Schwinger
model in \cite{HAG}, the model though did not loose its
solvability nevertheless unitarity failed to be maintained and the
model remained non attractive over a long period of time because
of its inability to provide the necessary condition for being
physically sensible and was left un studied till this severe
non-unitary problem  was removed in \cite{JR}. So a question may
automatically be arise whether the chiral generalization of
Thirring-Wess model face the same problem or it is free from that
severe un-physical situation? To explore it let us begin our
analysis in the following section on the model proposed here.
\section{Determination of theoretical  spectrum through lagrangian formulation}
A careful look reveals that the
 the  electromagnetic current related to lagrangian density (\ref{PLAG})
that takes part in the  interaction with matter field is
\begin{equation}
J^\mu= e(g^{\mu\nu}
 - \epsilon^{\mu\nu)}\partial_\nu\phi + m^2 A^\mu.
\end{equation}
Note that  $\partial_\mu J^\mu \ne 0$. So unlike the Schwinger
model the current does not conserve. Let us now proceed to study
the model through the standard lagrangian formulation.  The
Euler-Lagrange equations for the fields $A_\mu$ and $\phi$ that
flow from the same lagrangian density (\ref{PLAG}) are
\begin{equation}
\partial_\mu F^{\mu\nu} + e(g^{\mu\nu}-
\epsilon^{\mu\nu})\partial_\nu \phi + m^2 A^\mu = 0, \label{EL1}
\end{equation}
\begin{equation}
\Box\phi + e(g^{\mu\nu}- \epsilon^{\mu\nu})\partial_\nu A_\mu
=0.\label{EL2}
\end{equation}
Solving the above equations one may obtain the free field
solutions corresponding to the model (\ref{PLAG}) to which we now
turn. From equation (\ref{EL2}) we can write
\begin{equation}
\phi= -e(g^{\mu\nu}- \epsilon^{\mu\nu})\frac{\partial_\mu A_\nu
}{\Box} \label{PHI}. \end{equation}
 With the use of equation
(\ref{PHI}), equation (\ref{EL1}) turns into
\begin{equation}
\partial_\mu F^{\mu\nu} + e^2(g^{\nu\lambda}-
\epsilon^{\nu\lambda})\frac{\partial_\lambda
\partial_\kappa}{\Box}(g^{\kappa\mu}-
\epsilon^{\kappa\mu})A_\mu + m^2 A^\nu = 0. \label{FEQA}
\end{equation}
It can be shown that the above equations (\ref{EL1}) and
(\ref{EL2}) will be satisfied for the following most general form
of the field  $A_\mu$:
\begin{equation}
A^\mu=-\frac{e}{m^2}(\partial^\mu \phi +
\frac{(e^2-m^2)}{e^2}\epsilon^{\mu\nu}\partial_\nu \phi
-\frac{m^2}{e^2}\epsilon^{\mu\nu}\partial_\nu \chi), \label{EA}
\end{equation}

 Let us see how does it take place? From equation (\ref{EA}) it can be shown that
\begin{equation}
g^{\mu\nu}\partial_\mu A_\nu = -\frac{e}{m^2}\Box \phi
\label{CL0},
\end{equation}
and
\begin{equation}
\epsilon^{\mu\nu}\partial_\mu A_\nu =
-\frac{e}{m^2}[\frac{(e^2-m^2)}{e^2}\Box \phi -
\frac{m^2}{e^2}\Box \chi], \label{CL1}
\end{equation}
We, therefore, find that equations (\ref{CL0}) and (\ref{CL1}) are
capable of reproducing the equation (\ref{EL2}) as  well as
equation (\ref{PHI}),  if the field $\chi$ satisfies equation
$\Box \chi = 0$.  Using the expression of $A_\mu$ given in
(\ref{EA}) a straightforward calculation shows that  $A_\mu$ and
$\phi$ are related by the following second order differential
equations.
\begin{equation}
\Box A^\mu = -\frac{e}{m^2}(\partial^\mu \Box\phi +
\frac{(e^2-m^2)}{e^2}\epsilon^{\mu\nu}\partial_\nu \Box\phi),
\label{PA1}
\end{equation}
and
\begin{equation}
\partial^\mu \partial_\nu A^\nu =  -\frac{e}{m^2}\Box\partial^\mu
\phi. \label{PA2}
\end{equation}
Substituting the equations (\ref{PA1}) and (\ref{PA2}) into the
equation (\ref{EL1}) we find that the left hand side of the
equation (\ref{EL1})  turns into the following
\begin{equation}
\partial_\mu F^{\mu\nu} + e(g^{\mu\nu}-
\epsilon^{\mu\nu})\partial_\nu \phi + m^2 A^\mu = \Box A^\mu -
\partial^\mu \partial_\nu A^\nu + e(g^{\mu\nu}-
\epsilon^{\mu\nu})\partial_\nu \phi + m^2 A^\mu. \label{EPX}
\end{equation}
If we now substitute $A_\mu$ in (\ref{EPX}), the right hand side
of the equation (\ref{EPX}) turns into $\frac{e^2 -
m^2}{em^2}(\Box + \frac{m^4}{m^2 - e^2})\Box(\phi + \chi)$. So in
order to satisfy equation (\ref{EL1}) the field combination $(\phi
+ \chi)$ must satisfy the equation
\begin{equation}
(\Box + \frac{m^4}{m^2 - e^2})\Box(\phi + \chi) = 0.
\end{equation}
So, it is clearly seen that with the definition $A_\mu$ given in
(\ref{EA}), the Euler-Lagrange equations (\ref{EL1}) and
(\ref{EL2}) followed from the lagrangian (\ref{PLAG}), get
satisfied when the following two conditions hold.
\begin{equation}
(\Box + \frac{m^4}{m^2 - e^2})\Box(\phi + \chi) = 0,\label{SPEC01}
\end{equation}
\begin{equation}
\Box \chi = 0. \label{SPEC02}
\end{equation}
In fact, these are the theoretical spectra in the Lagrangian
formulation. The equation (\ref{SPEC01}), describes a free massive
boson with square of the mass $\tilde{m}^2 = \frac{m^4}{m^2 -
e^2}$ and equation (\ref{SPEC01}) represents a massless boson.
Therefore, the above computation over the proposed model in the
lagrangian formulation suggests that the system under
consideration contains two free fields $\Box(\phi + \chi)$ and
$\chi$. It also indicates that one of these two fields $\Box(\phi
+ \chi)$ carries a mass with square of the mass $\tilde{m}^2 =
\frac{m^4}{m^2 - e^2}$ but the other one $\chi$ is of vanishing
mass.

 Though the equations of the spectrum  (\ref{SPEC01}) and
(\ref{SPEC02}), come out as a condition to satisfy equation
(\ref{EL1}) and (\ref{EL2}), it does not mean that these two do
not have any physical significance. We will be able to see a
meaningful correspondence of these two with the spectrum that
would be obtained in the Hamiltonian formulation which we are
going to consider in the next section. Before that let us see what
does the combination of the fields $(\phi + \chi)$ stands for. To
see it let us compute the expression for
$\epsilon_{\mu\nu}\partial^\nu A^\mu$. We find that in $(1+1)$
dimension
\begin{equation}
\pi_1= -\epsilon_{\mu\nu}\partial^\nu A^\mu = \frac{m^2-e^2}{m^2e}
\Box(\phi + \chi). \label{PI}
\end{equation}
Here $\pi_1= \dot A_1 -A_0'$,  the momenta corresponding to the
field $A_1$. The condition $\Box \chi =0$, is also required to get
the relation (\ref{PI}).  The equation (\ref{PI}) will be useful
to see the correspondence between the spectra coming out here from
lagrangian formulation to the spectrum that would come out from
the hamiltonian formulation which we have described in the last
paragraph of the next section. We will now proceed to carry out
the hamiltonian formulation of the system.

\section{Constraint analysis and determination of the theoretical spectrum}
 To study the model in the hamiltonian formulation, the standard quantization of
constrained system due to Dirac has been  employed here. This
formulation not only helps to get  the theoretical spectrum but
also it enables to understand the positive definiteness of the
energy of this system  during the course of its analysis. Needless
to mention that positive definiteness is the fist step of ensuring
the unitary property of a field theoretical model. To investigate
the fate of this model in the present scenario we thus proceed to
quantize the theory.

Applications of Dirac's formalism towards the quantization of the
proposed theory involves the identification of the constraints of
 the same that remains embedded within its
 phase space. We, therefore, proceed to identify the constraint of the theory
at the early stage of our analysis.
 To this end, we require to calculate the canonical momenta of the fields with
  which the model is constituted.
 The momentum corresponding to the field $\phi$,
 $A_0$ and $A_1$ respectively are
 \begin{equation}
 \pi_\phi = \dot\phi + e(A_0 - A_1), \label{MO1}
 \end{equation}
 \begin{equation}
 \pi_0 = 0,\label{MO2}
 \end{equation}
 \begin{equation}
 \pi_1 = \dot A_1 - A_0'.\label{MO3}
 \end{equation}
 The above three equations (\ref{MO1}), (\ref{MO2}) and (\ref{MO3}), help us to obtain
 the hamiltonian
 through the Legendre transformation
 \begin{equation}
 H_B = \int dx [\pi_\phi\dot\phi + \pi_1{\dot A}_1 + \pi_0{\dot
 A}_0 - {\cal L}], \end{equation} which ultimately gives the following
 hamiltonian density for the system we are interested in.
 \begin{eqnarray}
{\cal H}_B = \frac{1}{2} (\pi_1^2 + \pi_\phi^2 +\phi'^2) +
\pi_1A_0'- e(\pi_\phi + \phi') (A_0 -A_1)+ \frac{1}{2}e^2(A_0 -
A_1)^2 - \frac{1}{2}m^2(A_0^2 - A_1^2). \label{HAM0}
\end{eqnarray}
 Equation (\ref{MO2}), is independent of $\dot A_0$. So it
 is the primary constraint of the theory. According to Dirac's
 prescription \cite{DIR},
 the further analysis from this stage would have to be done using the
  effective hamiltonian in
 stead of using the canonical hamiltonian obtained directly from the Legendre
 transformation in equation (\ref{HAM0}), and the effective hamiltonian in this
  situation is
 \begin{equation}
 H_{eff}= H + \int dx u_0 \pi_0.
 \end{equation}
 The Lagrangian multiplier (velocity) $u_0$ is yet to be determined. It will be fixed later.
 From the point of view of the physical consistency it stands as an
 essential requirement that the primary
 constraint is constrained to satisfy
 the condition $\dot\pi_0=[H(x), \pi_0(y)] \approx 0$, because it has
  to be preserved for all time  and that leads to the
  secondary constraint
 \begin{equation}
 G = \pi_1' + 2e(\pi_\phi + \phi') +(m^2 -e^2)A_0 +  e^2 A_1\approx 0.
 \label{GAUS0}
 \end{equation}
 It is known as the Gauss law of the theory. This constraint (\ref{GAUS0}), needs
 to preserve like the primary constraint (\ref{MO2}. However, preservation of the constraint
 (\ref{GAUS0}),
 does not give rise to any new
 constraint. Instead, it fixes the velocity $u_0$. This indicates that these two
 second class
 constraints are embedded within the phase space of the system and these two
 stand as the initial input to calculate the Dirac brackets.

According to the Dirac terminology \cite{DIR}, the constraints
(\ref{MO2}) and (\ref{GAUS0}) both are weak conditions up to this
stage. If it is now attempted to impose these into the hamiltonian
treating these two as strong
  condition, the hamiltonian will be then be  reduced to
 \begin{eqnarray}
 H_R &=& \int dx[\frac{1}{2}\pi_1^2 + \frac{1}{2}\frac{1}{m^2-e^2}
\pi_1'^2+ \frac{1}{2}\frac{m^2}{m^2-e^2}(\pi_\phi^2 +
 \phi^2)+ \frac{e^2}{m^2 -e^2}
 \pi_1'A_1 \nonumber \\
&+&\frac{e^2}{m^2-e^2}\pi_\phi\phi' + \frac{e}{m^2
-e^2}\pi_1'\phi'
 +  \frac{e}{m^2 -e^2}\pi_1'\phi' +
  \frac{em^2}{m^2-e^2}A_1\pi_\phi \nonumber \\
   &+& \frac{em^2}{m^2-e^2}A_1\phi' +
  \frac{1}{2}\frac{m^4}{m^2-e^2}A_1^2] .
 \label{RHA}
 \end{eqnarray}
 But the price that has to be paid for this is to replace  the  canonical Poission brackets
 by the corresponding Dirac bracket \cite{DIR} because the  Poission brackets become inadequate
 when the constraints are plugged in strongly into the Hamiltonian. It
 is known that Dirac bracket
 between the two variables A(x) and B(y) is defined by
 \begin{equation}
 [A(x), B(y)]^* = [A(x), B(y)] - \int[A(x) \omega_i(\eta)]
 C^{-1}_{ij}(\eta, z)[\omega_j(z), B(y)]d\eta dz, \label{DEFD}
 \end{equation}
 where $C^{-1}_{ij}(x,y)$ is given by
 \begin{equation}
 \int C^{-1}_{ij}(x,z) [\omega_i(z), \omega_j(y)]dz = 1.
 \label{INV} \end{equation}
 Here $\omega
 _i$'s represents the second class constraints that remains embedded within the phase space
 of the theory.
 The matrix $C^{-1}(x,y)$ for the present theory is given by
\begin{equation}
C^{-1}(x, y) =
\frac{1}{m^2-e^2} \left(\begin{array}{cc} 0 & \delta(x-y)\\
                -\delta(x-y) & 0 \end{array}\right),
 \label{MAT}\end{equation}
With the help of equation (\ref{DEFD}), the Dirac brackets among
the fields $A_1$,
 $\pi_1$, $\phi$, and $\pi_\phi$ are calculated:
 \begin{equation}
 [A_1(x), A_1(y)]^* =0
 \label{DR1}
 \end{equation}
\begin{equation}
 [A_1(x) ,  \pi_1(y)]^* =\delta(x
 -y)\label{DR2}
 \end{equation}
 \begin{equation}
 [\phi(x) , \phi(y)]^* =\delta(x -y)\label{DR3}
 \end{equation}
\begin{equation}
 [\phi(x) , \pi_\phi(y)]^* =0 \label{DR4}
 \end{equation}
Note that the above equations imply that the Dirac brackets retain
its own Poission bracket structures here which was not the case
when it was studied in \cite{THAR} with the Faddeevian class of
regularization.
 Making use of the Dirac brackets (\ref{DR1}), (\ref{DR2}), (\ref{DR3})
 and (\ref{DR4}), the equations of motion for the
 fields with which the hamiltonian (\ref{RHA}) is constituted  are
 computed and it is found that the following first order equations of motion
 are resulted in
\begin{equation}
 \dot A_1 = \pi_1 - \frac{e^2}{m^2 - e^2}A_1'- \frac{e}{m^2 - e^2} \pi_\phi'
  - \frac{e}{m^2 - e^2}\phi'' - \frac{1}{m^2 - e^2} \pi_1'' \label{EQM1}
 \end{equation}
 \begin{equation}
 \dot \pi_1 = - \frac{e^2}{m^2 - e^2}\pi_1' -\frac{em^2}{m^2 - e^2}\pi_\phi
 - \frac{em^2}{m^2 - e^2}\phi' - \frac{m^4}{m^2 - e^2}A_1\label{EQM2}
 \end{equation}
 \begin{equation}
 \dot\phi = \frac{m^2}{m^2 - e^2}\pi_\phi+ \frac{e^2}{m^2 - e^2}\phi'
 + \frac{e}{m^2 - e^2}\pi_1' + \frac{em^2}{m^2 - e^2}A_1 \label{EQM3}
 \end{equation}
\begin{equation}
 \dot\pi_\phi = \frac{m^2}{m^2 - e^2}\phi'' + \frac{e^2}{m^2 - e^2}\pi_\phi'
+ \frac{e}{m^2 - e^2}\pi_1''+ \frac{em^2}{m^2 -
e^2}A_1'\label{EQM4}
 \end{equation}
The above first order equations of motion get simplified into the
following second order differential equations after a little
algebra.
 \begin{equation}
 (\Box + \frac{m^4}{m^2 - e^2})\pi_1 = 0\label{SPEC0},
 \end{equation}
 \begin{equation}
(\Box + \frac{m^4}{m^2 - e^2})(A_1+ \frac{e}{m^2}\phi) = 0,
\label{SPEC1}
 \end{equation}
\begin{equation}
\Box(\phi+ \frac{e}{m^2}\pi_1) = 0,\label{SPEC2}
\end{equation}
\begin{equation}
\Box(\pi_\phi+ \frac{e}{m^2}\pi_1') = 0 \label{SPEC3}
\end{equation}
The above four equations (\ref{SPEC0}), (\ref{SPEC1}),
(\ref{SPEC2}) and (\ref{SPEC3}) suggest that the field $A_1+
\frac{e}{m^2}\phi$ describe a massive boson with square of the
mass $\tilde{m}^2 = \frac{m^4}{m^2 - e^2}$ and the field $\phi+
\frac{e}{m^2}\pi_1$ represents a boson with vanishing mass. The
field $\pi_1 $ and $\pi_\phi+ \frac{e}{m^2}\pi_1'$ may be
considered as the momenta corresponding to the field $A_1+
\frac{e}{m^2}\phi$ and $\pi_\phi+ \frac{e}{m^2}\pi_1'$
respectively because the pair of  fields describing equations
\ref{SPEC0}) and (\ref{SPEC1}) satisfy canonical poisson bracket
among themselves and the pairs describing the equations
(\ref{SPEC2}) and (\ref{SPEC3}) also satisfy the canonical poisson
bracket. Note that $m^2$ must be greater than $e^2$ in order to
get the mass of the massive boson a physical (positive) one. The
condition $m^2 > e^2$ is essential in order the Hamiltonian
(\ref{RHA}) to be positive definite which is the basic footing of
ensuring uniatrity of the theory. One more thing which we would
like to address here is that some practical benefit may also be
followed from this model. The massless bosons found in the
spectrum is equivalent to a massless fermion in two dimensions so
it can be taught of as fermion in a de-confined state \cite{JR,
PM, MG, AR}.

We will now be able to see the correspondence between the spectrum
obtained here and the spectrum that came out in lagrangian
formulation.  If we look carefully towards the equations
(\ref{SPEC01}), (\ref{PI}) and (\ref{SPEC0}) we will be able to
see that the equations (\ref{SPEC01}) and (\ref{SPEC0}) are
identical. Using equation (\ref{PI}) and (\ref{SPEC0}) it can be
ensured that the massless field $\chi$ that appeared in the
lagrangian formulation is nothing but the field $\phi+
\frac{e}{m^2}\pi_1$. So the spectrum that came out as a condition
in the lagrangian formulation does not fail to carry proper
physical meaning.

\section{Inclusion of counter term containing the ambiguity parameters}
In Sec. II, Chiral Thirring-Wess model has been defined by the
 following generating functional
 \begin{equation}
 Z[A] = \int d\psi d\bar\psi e^{\int d^2x{\cal L}_f}
 \end{equation}
 where the definition of ${\cal L}_f$ is available from the
 equation (\ref{LF}).
As mentioned in Sec. II, the dynamics of the $A_\mu$ field is
governed by he Proca field
 and the lagrangian of which is given in equation (\ref{PROCA})
In Sec. II, it has  already been mentioned that in the lagrangian
(\ref{LF}), right handed fermion remains uncoupled
 when vector interaction is replaced by the  chiral interaction. So integration
 over this right
 handed part leads to field independent counter part which can be
 absorbed within the normalization. However the integration over the left handed
 fermion is much involved because one needs to regularize  the
 fermionic determinant during the process of integration since
 the determinant has a diverging
 nature,
 and after a careful calculation
  one arrives at the following generating functional \cite{PM, SUBIR0, SUBIR, THAR}
 \begin{eqnarray}
 Z[A] &=& \int d\psi_L d\bar\psi_L\bar\psi_L
 \gamma^\mu(i\partial_\mu + 2e\sqrt\pi
 A_\mu)\psi_L  \nonumber \\
 &=&exp\frac{ie^2}{2}\int d^2x A_\mu[M_{\mu\nu} - (\partial^\mu
 +\tilde\partial^\mu) \frac{1}{\Box} (\partial^\nu +
 \tilde\partial^\nu)]A_\nu, \label{GEN1}
 \end{eqnarray}
 In general, the elements of the
 $M_{\mu\nu}$ can take any arbitrary
 values. However, the model looses both its solvability and Lorentz
 invariance in that situation \cite{THAR}. In \cite{THAR}, we considered
  a symmetric form of
 $M_{\mu\nu}$:
\begin{equation}
 M_{\mu\nu} = \left(\begin{array}{cc} \tilde{a}  \quad \alpha \\
 \alpha  \quad  \gamma \\
 \end{array}\right)\delta(x-y).
\end{equation}
where regularization ambiguity got
 involved within
 the parameters $\tilde{a}$, $\alpha$ and $\gamma$. These parameters entered there in order to remove the
 divergence of the fermionic determinant since the evaluation of the determinant needs
  a {\it one loop correction}
 \cite{PM, SUBIR0, SUBIR}. It was found in \cite{THAR}  that all the parameters
   did not remain independent with each other. Some constrained among the ambiguity parameters
    were found to
   be essential
    for the model to be physically
sensible.

This generating functional (\ref{GEN}), when written there in
terms of the auxiliary
 field $\phi(x)$ it turned out to the following
 \begin{equation}
 Z[A] = \int d\phi e^{i\int d^2x {\cal L}_B},
 \end{equation}
 with
\begin{equation}
 {\cal L}_B
 =\frac{1}{2}(\dot\phi^2 - \phi'^2) + e(\dot\phi + \phi')(A_0
 - A_1) + \frac{1}{2}e^2(\tilde{a}A_0^2 + 2\alpha A_0A_1 + \gamma A_1^2).
 \end{equation}
 So the total lagrangian density with which we dealt there was
 \begin{eqnarray}
 {\cal L} =
 = \frac{1}{2}(\dot\phi^2 - \phi'^2) + e(\dot\phi + \phi')(A_0
 - A_1) + \frac{1}{2} ({\dot A_1}^2-{A'_0}^2) + {\cal L}_{mass}
 \label{PLAG0}\end{eqnarray}
where the term ${\cal L}_{mass}$ was
 \begin{equation}
 {\cal L}_{mass}=
 =\frac{1}{2}e^2[(\tilde{a} + \frac{m^2}{e^2})A_0^2 + 2\alpha A_0A_1 +
 (\gamma - \frac{m^2}{e^2})A_1^2].
 \end{equation}
 This Lagrangian in general failed to provide Poincar\'e invariant equations
 of motion. Ambiguity in the regularization allowed us to set two
 conditions $\tilde{a}+\frac{m^2}{e^2}=1$ and $m^2=e^2(1+\gamma-2\alpha)$
 without violating
 any physical principle. In our previous work \cite{THAR},  we showed that the condition
 $\tilde{a}+\frac{m^2}{e^2}=1$
 helped us to fit the model within the Faddeevian
 class \cite{FAD1, FAD2, SHATAS0, SHATAS} and the theory rendered an interesting Lorentz in variant
 theoretical spectrum though there was no lorentz covariance in the starting
 lagrangian provided the constraint $m^2=e^2(1+\gamma-2\alpha)$ among the ambiguity parameters
 were maintained.
Indeed, some other choices may also lead to some other physically
sensible theories.  We are free to
 choose Jjackiw-Rajaraman type of regularization for this model. In that case we have to choose
  the matrix
  $M_{\mu\nu}$  as
 $M_{\mu\nu} = ag_{\mu\nu}$. It keeps the model Lorentz
 covariant to start with and the lagrangian density for this
 situation  reads
\begin{eqnarray}
{\cal L}=
 = \frac{1}{2}(\dot\phi^2 - \phi'^2) + e(\dot\phi + \phi')(A_0
 - A_1)
 + \frac{1}{2} ({\dot A_1}^2-{A'_0}^2) + \frac{1}{2}e^2(a+
 \frac{m^2}{e^2})(A_0^2 -A_1^2)
 \label{PLAJR}\end{eqnarray}
 Note that the only difference between the model given in equation (\ref{PLAJR}) and the model
 considered in equation (\ref{PLAG}),  lies in
  is in the masslike terms for gauge fields.  To be more precise the difference between
  ambiguity (one loop correction)
   free situation and
  the situation where ambiguity (one loop correction) is taken into consideration lies in
  the masslike terms for gauge fields.
The masslike term
 in the lagrangian (\ref{PLAJR}) now turns into
 $\frac{1}{2}e^2(a+
 \frac{m^2}{e^2})(A_0^2 -A_1^2)$. Note that  the constant $m^2$ of the
 lagrangian (\ref{PLAG}) is now
 shifted to $e^2(a+\frac{m^2}{e^2})$. All the other terms remains unaltered. So the
 analysis will also follow the same direction as it has been made
 in Sec. II. and Sec. III.  Unlike the situation studied in \cite{THAR}, the constraint
 structure will also  remain the same except the shifting of
 the constant parameter $m^2$ to $e^2(a+ \frac{m^2}{e^2})$. A
 careful look will help to
 understand without much difficulty that there will be a shift in the mass of the massive
 field  because of this shifting of the constant parameter ${m}^2$ in the
  masslike term of (\ref{PLAJR}). Without going through the detail
  analysis it can be inferred
  that the square of the mass of the massive boson
 that will follow from the lagrangian (\ref{PLAJR})
 will be none other than
  $\tilde {m}^2=
 \frac{m^4 +a^2e^4+ 2a m^2e^2}{m^2 + e^2(a -1)}$. However, the  massless
 excitation will remain unaffected and which can be thought of as a
 de-confined fermion in $(1+1)$ dimension as usual.

\section{Conclusion}
 In this paper we have studied the Thirring-Wess model replacing its vector
 type interaction by the chiral type. The model has been analyzed both
 in the lagrangian and hamiltonian formulation. Drawing Euler-
 Lagrange equations into service
 we have obtained that the model bears a rich Lorentz invariant
 theoretical spectrum. The
 use of the standard method of quantization of constrained system by Dirac \cite{DIR}
 too confirms the same in much more transparent manner. The spectrum contains
  a massive boson like
the usual vector Schwinger model. A mass less boson has also been
found to appear like the chiral Schwinger model. One more
information which the hamiltonian of the present model provides
for is that its positive definiteness is  maintained for the
restriction $m^2 > e^2$. It is indeed  the basic requirement of a
theory to be sound in true physical sense. Not only that but also
it is  the first step of ensuring the unitarity of the theory. The
model however does not possess gauge symmetry. It is not unnatural
since its ancestor, i.e., Thiring-Wess model is not gauge
symmetric like the vector Schwinger model.

 The point on which we would like to emphasize is
that the chiral generation studied here does not need any counter
term in order to make it physically sensible like the chiral
generation of the Schwinger model made in \cite {HAG}. It is found
from our analysis that unlike the  chiral generation of Schwinger
model \cite{HAG} this model does not fail to maintain unitarity
and exactly solvable nature in the absence of  any counter term
containing ambiguity parameter,  and as a result, no ambiguity
parameter entered into the mass of the massive boson. In this
sense, Chiral generation of the Thiring-Wess model scores over the
chiral generation of Schwinger model \cite{HAG} since it carries
less complication, and this would not suffer from any un-physical
situation like the chiral generation of Schwinger model
\cite{HAG}.

It has already been mentioned that the gauge symmetry is absent
here like its ancestor. Though the absence of gauge symmetry does
not violate any physical principle nevertheless one may argue that
a gauge symmetric theory is advantageous because it reflects an
increased symmetry of the lagrangian, albeit it has to be
remembered that what is increased is not a physical symmetry of
the states but only a symmetry of the effective action which has
to be broken by gauge fixing. Even though, if we allow some
respect or emphasis  on the gauge symmetry of the effective action
that can also be met here with out much difficulty. There is
certainly room for
 converting this gauge non-invariant model into a structure where
 gauge symmetry is maintained
 and that too within the
 ambit of its own physical phase space. The method of converting a gauge
 non-invariant model into a gauge invariant one as suggested by
 of Mitra  and Rajaraman \cite{MR1, MR2} may be useful in this context. The method of
 bringing back gauge symmetry of a gauge non-invariant model in the extend
phase spacer is also well known and can be applied here as well.
However, one needs to introduce the appropriate Wess-Zumino term
\cite{WESS} in that case. In that situation some extra fields will
enter within the effective action. But it is not a matter to get
worried because the extra fields allocate themselves within the
un-physical sector of the theory.

The hamiltonian of the model under present consideration is
positive definite that gives the signature of maintenance of
unitarity. However, the the formal proof of
 certainly comes from the BRST quantization and there is no difficulty to
 carry it out for this model since it provides positive definite hamiltonian.
 Therefore, this model is able to satisfy all the important criteria in order
 to be granted as physically sensible in
 the lower dimensional field theoretical regime from whatever aspects
 it have been looked for.

 \noindent {\bf Acknowledgement}
 I would like to thank the Director of
 Saha Institute of Nuclear Physics, Calcutta, and the head of the theory
 group of the same institute for providing library and computer facilities
 of the Institute.
 
 \end{document}